\documentclass[11pt, a4paper,english,superscriptaddress,aps]{revtex4}
\usepackage{amsmath,amssymb,graphicx}
\makeatletter
\usepackage{babel}
\usepackage[active]{srcltx}
\usepackage{graphicx,color}

\usepackage{hyperref}

\newcommand{\be}{\begin{equation}}
\newcommand{\ee}{\end{equation}}
\newcommand{\bea}{\begin{eqnarray}}
\newcommand{\eea}{\end{eqnarray}}

\begin{document}

\title{On the causality aspects of the dynamical Chern-Simons modified gravity}

\author{P. J. Porf\'{i}rio, J. B. Fonseca-Neto, J. R. Nascimento, A. Yu. Petrov}
\affiliation{Departamento de F\'{\i}sica, Universidade Federal da Para\'{\i}ba\\
 Caixa Postal 5008, 58051-970, Jo\~ao Pessoa, Para\'{\i}ba, Brazil}
\email{jfonseca,jroberto,petrov@fisica.ufpb.br}

\begin{abstract}
We discuss the G\"{o}del-type solutions within the dynamical Chern-Simons modified gravity in four dimensions. Within our study, we show that in the vacuum case the causal solutions are possible which cannot take place within the non-dynamical framework. Another our result consists in the possibility for completely causal solutions for all types of matters we study in the paper, that is, relativistic fluid, cosmological constant, scalar and electromagnetic fields.
\end{abstract}

\maketitle

\section{Introduction}

Studies of different modified gravity models certainly represent themselves as one of the main lines of research in the modern theoretical physics. There are two main reasons of interest to these studies -- first, the need for a consistent quantum description of gravity caused by the well-known fact that the Einstein gravity is non-renormalizable  \cite{Veltman}, and hence implying a search for a some more generic gravity theory reducing to a general relativity in a specific limit, second, the discovery of the cosmic acceleration \cite{Riess} which called an interest for searching the theories allowing for its better explanation.

One of possible and most interesting extensions of the Einstein gravity is the four-dimensional Chern-Simons (CS) modified gravity allowing to implement the CPT and Lorentz symmetry breaking within a gravity context. Another essential feature of this theory consists in the fact that it naturally involves higher (third) derivatives of the metric fluctuation.  Originally, the CS extension for the gravity has been proposed in three-dimensional space-time in \cite{DJT} where it was shown to be completely consistent with unitarity and causality.  Its generalization for the four-dimensional case has been carried out in \cite{AlvWitten} where it was shown that in this case the gravitational CS term breaks the parity and is described by the so-called Chern-Simons coefficient, that is, the pseudoscalar field $\phi(x)$. In the same paper, this term was shown to be related with gravitational anomalies (further, it was shown that the relation between the four-dimensional gravitational CS term and the gravitational anomalies is similar to the relation between the Carroll-Field-Jackiw term and the Adler-Bell-Jackiw anomaly taking place for the vector field, see f.e. \cite{JackAmb}).

 Further, the interest to the gravitational CS term strongly increased due to the famous paper \cite{JaPi} which established a relation between the four-dimensional gravitational Chern-Simons term and the Lorentz symmetry breaking. for a special form of the CS coefficient chosen to be proportional to the constant vector (that is, $\phi(x)=k_\mu x^\mu$).  Clearly, it allows to consider the four-dimensional gravitational CS term an ingredient of the most generic Lorentz-breaking extension of the standard model \cite{Kostel}. Also, it calls the interest to study the four-dimensional CS modified gravity representing itself as the simplest Lorentz-breaking gravity model where the new additive term looks like a small correction to the general relativity.  The next step for study of the CS extension gravity has been done in \cite{Grumiller} where the CS coefficient $\phi(x)$  was suggested to possess a nontrivial dynamics satisfying its own equations of motion, with the new theory has been called the dynamical Chern-Simons modified gravity (DCSMG) (for a review on DCSMG, see also \cite{Yunes}).

From the purely gravitational viewpoint, the consistency of any extension of the general relativity should be verified through study of the behavior of known solutions of general relativity within the new theory. Already in \cite{Grumiller} this consistency has been demonstrated for metrics possessing spherical or axial symmetry. At the same time, some known solutions of the general relativity fail to be consistent within the DCSMG. The paradigmatic example is the Kerr metric which should be modified by the terms proportional to some terms proportional to $\phi(x)$ to achieve the lower-order consistency \cite{Konno}, therefore, in general, the consistency of known general relativity solutions within the DCSMG is an open problem.

One of the most interesting classes of gravitational solutions whose consistency within the DCSMG clearly should be verified are those ones breaking the causality, or, as is the same, allowing for the closed time-like curves (CTCs). It is believed that CTCs are eliminated by a consistent theory of quantum gravity, since quantum effects would prevent pathological solutions of GR, but this is a conjecture \cite{Hawking} (the possibility to avoid the CTCs for specific metrics has been discussed in different papers, for example, it was noted in \cite{Deser} for the cosmic string space). An observer traveling along CTCs can return in time, i.e., such an observer can visit its own past generating a time travel and causality paradoxes. Various examples of GR solutions containing CTCs are known, for example, Van Stockum metric \cite{VS} and Gott time machine \cite{Gott}.  But, the most famous known example of such a solutions is the G\"{o}del metric \cite{Godel}. Further, this metric has been generalized  in \cite{Reb}, where a number of aspects of the new metric has been studied. In particular, it was showed in these papers that in certain cases the CTCs are ruled out for some special relations between constant parameters of this metric.  Therefore, it is interesting to verify the consistency of the G\"{o}del-type metric proposed in \cite{Reb} within the DCSMG, especially, to obtain the conditions for parameters allowing to rule out the CTCs. Some preliminary studies in this direction have been performed in \cite{ourGodel}, where, however, only the case of the original G\"{o}del metric has been considered, and only the case of the vanishing Cotton tensor was considered, that is, the left-hand side of the equations of motion is not modified. Therefore, it is interesting to consider the compatibility of G\"{o}del-type metric within the DCSMG, with no restrictions on the Cotton tensor. Thus, our study will be the natural continuation of our previous paper \cite{PP} where the G\"{o}del-type metric was considered within the non-dynamical CS modified gravity.

The structure of our paper looks like follows. In the section 2, we give a review of the G\"{o}del-type metric. In the section 3, we discuss the general features of the DCSMG. The section 4 is the main part of our paper where we solve the equations of motion of our theory and discuss the properties of solutions. In the section 5, we discuss our results.

\section{G\"{o}del-type metrics}

This section is a review of the G\"{o}del-type metrics and its causality properties necessary for further purposes.

One of first causality violating solutions of Einstein equations was obtained by G\"{o}del \cite{Godel}. The rotating G\"{o}del universe is characterized by following line element:
\begin{equation}
ds^2=-[dt+H(x)dy]^2+D^{2}(x)dy^{2}+dx^{2}+dz^{2},
\label{Godel}
\end{equation}
where the functions $H(x)$ and $D(x)$ look like
\begin{equation}
H(x)=e^{mx},\quad D(x)=\frac{e^{mx}}{\sqrt{2}},
\end{equation}
and the parameters of solutions are related with the matter content, that is, the density $\rho$ and the cosmological term $\Lambda$ through the relations:
\begin{equation}
\begin{split}
& m^2=2\omega^2=\kappa\rho;\\
& \Lambda=-\kappa\frac{\rho}{2}
\end{split}
\end{equation}
where $\kappa$ is the Einstein constant and $\omega$ is the vorticity of matter. Furthermore, the G\"{o}del metric satisfies the conditions of the space-time homogeneity \cite{Reb}. The metric (\ref{Godel}) can be generalized in a manner proposed in \cite{Reb}:
 \begin{equation}
 ds^2=-[dt+H(r)d\theta]^2+D^{2}(r)d\theta^{2}+dr^{2}+dz^{2},
 \label{metric}
 \end{equation}
where the functions $H(r)$ and $D(r)$ must obey the conditions below to be ST-homogeneous,
\begin{equation}
\begin{split}
&\frac{H^{'}(r)}{D(r)}=2\omega,\\
&\frac{D^{''}(r)}{D(r)}=m^{2},
\end{split}
\label{ST}
\end{equation}
where the prime denotes the derivative with respect to $r$. The parameters $(m^2,\omega)$ are constants such that $\omega\neq0$ and $-\infty \leq m^2 \leq \infty$. From now we will consider only space-time homogeneous  G\"{o}del-type metrics. There exist three different classes of solutions for Eqs. (\ref{ST}) which can be distinguished by the sign of $m^2$:

i)\textit{hyperbolic class}: $m^2>0$, $\omega\neq 0$:
\begin{equation}
\begin{split}
&H(r)=\frac{2\omega}{m^2}[\cosh(mr)-1],\\
&D(r)=\frac{1}{m}\sinh(mr),\\
\end{split}
\end{equation}

$\,\,$

ii)\textit{trigonometric class}: $-\mu^2=m^2<0$, $\omega\neq 0$:
\begin{equation}
\begin{split}
&H(r)=\frac{2\omega}{\mu^2}[1-\cos(\mu r)],\\
&D(r)=\frac{1}{\mu}\sin(\mu r),\\
\end{split}
\label{trigo}
\end{equation}

$\,\,$

iii)\textit{linear class}: $m^2=0$, $\omega\neq 0$:
\begin{equation}
\begin{split}
&H(r)=\omega r^2,\\
&D(r)=r.\\
\end{split}
\label{linear}
\end{equation}

 The hyperbolic class presents as a particular case $m^2=2\omega^2$ that corresponds to the usual G\"{o}del metric (\ref{Godel}).

The causality properties of the G\"{o}del-type metric are described by theorems obtained in \cite{Reb}:
 
   \textbf{Theorem 1}: The necessary and sufficient conditions for a Riemannian G\"{o}del-type manifold to be locally homogeneous in space and time are those given by Eq.(\ref{ST}).
    
   \textbf{Theorem 2}: The Riemannian G\"{o}del-type manifolds homogeneous in space and time are characterized  by two parameters $(m^2,\omega)$. Equal pairs of $(m^2,\omega)$ specify locally equivalent metrics.

   \textbf{Theorem 3}: The Riemannian G\"{o}del-type manifolds homogeneous in space and time admit a group of  isometries such that: $G_{5}$ (when $\omega\neq 0$), $G_{6}$ (for $\omega=0$) and $G_{7}$ (for the class $m^2=4\omega^2$).
   
  Now, we can discuss in more details the causality violation. We rewrite the Eq. (\ref{metric})  in the form
\begin{equation}
   ds^2=-dt^2-2H(r)dt d\theta+G(r)d\theta^{2}+dr^{2}+dz^{2},
\end{equation}     
 where $G(r)=D^2 (r)-H^2 (r)$. The circle defined by $C=\lbrace(t,r,\theta,z); \, t=t_0, r=r_0, \theta \in [0, 2\pi], z=z_0\rbrace$, is a CTC if $G(r)$ becomes negative for a range of $r$-values ($r_1 <r<r_2$) \cite{Reb}. For $m^2=0$, the linear class, one non-causal region exists for $r>r_{c}$, where $r_{c}=1/\omega$ is the critical radius. 
  For $m^2=-\mu^2$, the trigonometric class, there is an infinite number of alternating causal and non-causal regions. The region, $0< m^2<4\omega^2$ corresponds to the hyperbolic class, so that there is one non-causal region for $r>r_{c}$, with the critical radius is given by
\begin{equation}
\sinh^2\bigg(\frac{m r_{c}}{2}\bigg)=\bigg(\frac{4\omega^2}{m^2}-1 \bigg).
\end{equation}
 Nevertheless,  the region $m^2\geq 4\omega^2$ corresponding to  hyperbolic class as well, is completely causal, so, in this case, there are no CTCs.

\section{Chern-Simons modified gravity}

The dynamical Chern-Simons (CS) modified gravity action \cite{Konno,Yunes} with the cosmological constant is described by the action
 \begin{equation}
 S_{CS}=\frac{1}{2\kappa}\int d^{4}x \left[\sqrt{-g}(R-2\Lambda)+\sqrt{-g}\bigg(\frac{1}{4}\phi\ ^{*}RR\bigg)-\beta\sqrt{-g}\bigg(\frac{1}{2}\nabla^{\mu}\phi\nabla_{\mu}\phi\bigg)\right]+S_{mat}, 
\label{CSMG} 
 \end{equation}
where  $\phi$ is a scalar field, $\beta$ is a dimensionful coupling constant and $S_{mat}$ is the matter action. In this action, the  $\phi$ has dimension of squared length, $\left[\phi\right]=L^2$ , thus one can see that the dimension of $\beta$ is $\left[\beta\right]=L^{-4}$. The term, $^{*}RR$, is the Pontryagin density whose integral on a closed manifold is a topological invariant. It is defined as  follows
 \begin{equation}
 \begin{split}
 \ ^{*}RR &\equiv \ ^{*}R^{\mu\ \ \gamma\sigma}_{\ \ \nu}R^{\nu}_{\ \mu\gamma\sigma}=\frac{1}{2}\varepsilon^{\gamma\sigma\tau\eta}R^{\mu}_{\ \ \nu\tau\eta}R^{\nu}_{\ \mu\gamma\sigma},\\
 \end{split}
 \end{equation}
 where $\varepsilon^{\gamma\sigma\tau\eta}$ is the Levi-Civita tensor. Moreover, the Pontryagin density is related with the usual four-dimensional gravitational Chern-Simons-like term as
 \begin{equation}
 ^{*}RR=2\nabla_{\alpha}\left[\varepsilon^{\alpha\beta\gamma\tau}\left(\Gamma^{\mu}_{\beta\nu}\partial_{\gamma}\Gamma^{\nu}_{\tau\mu}+\frac{2}{3}\Gamma^{\mu}_{\beta\nu}\Gamma^{\nu}_{\gamma\eta}\Gamma^{\eta}_{\tau\mu}\right)\right],
 \end{equation}
and the term in brackets is commonly called as CS topological current. Using this equation, we have
 \begin{equation}
\frac{1}{4}\int d^{4}x \sqrt{-g}\ \phi\ ^{*}RR=-\frac{1}{2}\int d^{4}x \sqrt{-g}\ v_{\alpha}\varepsilon^{\alpha\beta\gamma\tau}\left(\Gamma^{\mu}_{\beta\nu}\partial_{\gamma}\Gamma^{\nu}_{\tau\mu}+\frac{2}{3}\Gamma^{\mu}_{\beta\nu}\Gamma^{\nu}_{\gamma\eta}\Gamma^{\eta}_{\tau\mu}\right),
 \end{equation}
  where we integrate by parts and define $v_{\mu}=\nabla_{\mu}\phi$. The scalar field $\phi$ is called the CS coefficient or the CS coupling field and measures how the modified theory deforms the GR. The case where $\phi$ is a constant makes the CS term to be a total derivative and hence reduces to GR, therefore the non-trivial solutions of modified theory require the scalar field to be a non-trivial function of space-time
coordinates.

 Varying the CS modified action  with respect to $g_{\mu\nu}$ and $\phi$, we obtain the field equations, taking $\kappa=1$, respectively,
 \begin{eqnarray}
R_{\mu\nu}-\frac{1}{2}Rg_{\mu\nu}+C_{\mu\nu}+\Lambda g_{\mu\nu}&=&T_{\mu\nu}+T^{\Phi}_{\mu\nu};\nonumber\\
\beta\square\phi&=&-\frac{1}{4}\,^{*}RR,
\label{KG}
\end{eqnarray}
where  $\square\equiv g^{\mu\nu}\nabla_{\mu}\nabla_{\nu}$ denotes the d'Alembertian operator.

The energy-momentum tensor of the matter $T_{\mu\nu}$ is defined by
\begin{equation}
T^{\mu\nu}=-\frac{2}{\sqrt{-g}}\bigg(\frac{\delta\mathcal{L}_{m}}{\delta g_{\mu\nu}} \bigg),
\end{equation} 
and the contributions of the scalar field $\phi$ can be expressed in terms of  the energy-momentum tensor $T^{\phi}_{\mu\nu}$
\begin{equation}
T^{\phi}_{\mu\nu}=\beta\left[(\nabla_{\mu}\phi) (\nabla_{\nu}\phi)-\frac{1}{2}g_{\mu\nu}(\nabla_{\lambda}\phi) (\nabla^{\lambda}\phi)\right] ,
\end{equation}  
   and $C^{\mu\nu}$ is the Cotton tensor, symmetric and traceless, arising due to variation of the CS terms, the explicit expression for the Cotton tensor is,
 \begin{eqnarray}
 C^{\mu\nu}=-\frac{1}{2}[v_{\sigma}(\varepsilon^{\sigma\mu\alpha\beta}\nabla_{\alpha}R^{\nu}_{\beta}+
\varepsilon^{\sigma\nu\alpha\beta}\nabla_{\alpha}R^{\mu}_{\beta})+v_{\sigma\tau}({}^{*}R^{\tau\mu\sigma\nu}+\ ^{*}R^{\tau\nu\sigma\mu})],
\end{eqnarray}    
 where $v_{\sigma\tau}=\nabla_{\sigma}v_{\tau}$. The covariant divergence of the Cotton tensor is
 \begin{equation}
  \nabla_{\mu}C^{\mu\nu}=\frac{1}{8}v^{\nu}\ ^{*}RR.
\label{nabla}
\end{equation}  
The energy-momentum tensor of scalar field $\phi$ is non-vanishing, thus even for the vacuum the energy density associated at scalar field $\phi$ is still non-trivial.
  
\section{G\"{o}del-type metrics in dynamical framework}

Now, let us start with study of the G\"{o}del-type metrics.

For a sake of simplicity we adopt a local Lorentz tetrad, $w^{A}=e^{A}_{\,\,\,\mu}dx^\mu$. Thus, for the metric (\ref{metric}) we have
\begin{eqnarray}
w^{(0)}&=&dt+H(r)d\theta;\nonumber\\
w^{(1)}&=&dr;\nonumber\\
w^{(2)}&=&D(r)d\theta;\nonumber\\
w^{(3)}&=&dz.
\label{TB}
\end{eqnarray}

We use the Minkowski metric $\eta_{AB}=diag(-1,+1,+1,+1)$, and  the Riemann tensor and Ricci tensor are defined in frame (\ref{TB}), respectively, by
\begin{eqnarray}
R^{D}_{\,\,\,CAB}&=&\partial_{A}\,\omega^{D}_{\,\,\,B C}-\partial_{B}\,\omega^{D}_{\,\,\, A C}+\omega^{E}_{ \,\,\,BC}\,\omega^{D}_{\,\,\,AE}-\omega^{E}_{ \,\,\,AC}\,\omega^{D}_{\,\,\,BE}+2\Omega^{E}_{ \,\,\,BA}\,\omega^{D}_{\,\,\,EC},\\
R_{CB}&=&R^{D}_{\,\,\,CDB},\nonumber
\end{eqnarray} 
where we choose the following conventions $\Omega^{E}_{ \,\,\,BA}=e_{B}^{\,\,\,\nu}e_{A}^{\,\,\,\mu}\partial_{[\mu}e^{E}_{\,\,\,\nu]}$ are the anholonomy coefficients and $\omega^{B}_{\,\,\, AC}=e^{B}_{\,\nu}e^{\,\,\mu}_{C}e^{\,\,\alpha}_{A}\Gamma^{\nu}_{\,\,\mu\alpha}-e^{\,\,\mu}_{C}e^{\,\,\nu}_{A}\partial_{\mu}e^{B}_{\,\,\nu}$ are the connection coefficients in the frame (\ref{TB}), or Cartan connection, related to the Levi-Civita connection ($\Gamma^{\mu}_{\,\,\alpha\nu}$). We denote coordinate indices by small Greek letters, and tetrad indices by capital Latin letters.

 The non-vanishing components of  Ricci tensor are given by
  \begin{equation}
	\label{ricci2}  
	R_{(0)(0)}=2\omega^2, \qquad R_{(1)(1)}=R_{(2)(2)}=2\omega^2-m^2,
  \end{equation} 
 which implies in constant $R_{AB}$, as a consequence the $T_{AB}$ is diagonal and constant. 

  The non-vanishing components of  Cotton tensor $C_{AB}$ are:
  \begin{eqnarray}
  C_{(0)(0)}&=&2\bigg(\frac{\partial \phi}{\partial z}\bigg)\omega(4\omega^2-m^2),\nonumber\\
  C_{(0)(1)}&=&-\frac{1}{2}\bigg(\frac{\partial^2 \phi}{\partial z \partial t}H(r)-\frac{\partial^2 \phi}{\partial z \partial \theta}\bigg)\frac{(4\omega^2-m^2)}{D(r)},\nonumber\\
   C_{(0)(2)}&=&-\frac{1}{2}\frac{\partial^2 \phi}{\partial z \partial r}(4\omega^2-m^2);\nonumber\\
  C_{(0)(3)}&=&-\frac{\partial \phi}{\partial t}\omega(4\omega^2-m^2),\nonumber\\
 C_{(1)(1)}&=&\bigg(\frac{\partial \phi}{\partial z}\bigg)\omega(4\omega^2-m^2)
 \nonumber\\  
   C_{(1)(3)}&=&\frac{1}{2}\bigg(\frac{\partial^2 \phi}{ \partial t^2}H(r)-\frac{\partial^2 \phi}{\partial \theta \partial t}\bigg)\frac{(4\omega^2-m^2)}{D(r)};\nonumber\\
	C_{(2)(2)}&=&\bigg(\frac{\partial \phi}{\partial z}\bigg)\omega(4\omega^2-m^2)
 \nonumber\\  
    C_{(2)(3)}&=&\frac{1}{2}\frac{\partial^2 \phi}{\partial t \partial r}(4\omega^2-m^2).
  \label{CT}
  \end{eqnarray}
  where $\phi=\phi(t,r,\theta,z)$. The geometry $m^2=4\omega^2$ cancels all $C_{AB}$.
   
The Eqs. (\ref{KG}) take the form in the frame (\ref{TB}):   
 \begin{eqnarray}
  R_{AB}+C_{AB}&=&\bigg(T_{AB}-\frac{1}{2}T\eta_{AB}\bigg)+\bigg(T^{\phi}_{AB}-\frac{1}{2}T^{\phi}\eta_{AB}\bigg)+\Lambda\eta_{AB},\nonumber\\
\beta\square\phi&=&0,
\label{KGd}
\end{eqnarray}
where $T=T^{A}_{\,\,\,A}$ is the trace of energy-momentum tensor $T_{AB}$ and $T^{\phi}=\eta^{AB}T^{\phi}_{AB}$ is the trace of the energy-momentum tensor for the CS field. We note that the Pontryagin density vanishes for the G\"{o}del-type metric. Differently of the non-dynamical case, now the CS field is  treated as the dynamical one, thus, it satisfies the field equation given by Eq.(\ref{KGd}), which for G\"{o}del-type metrics is identical to the known Klein-Gordon equation for zero-mass scalar fields. We use the same matter sources considered as in \cite{Reb,PP}. 
  The next step is to solve the field equations in order to obtain a form for the CS field. For this, we use the facts that $T_{AB}$ and $R_{AB}$ are diagonal in the frame (\ref{TB}). As a consequence, the field equations for $A\neq B$ reduce to
\begin{equation}
C_{AB}=T^{\phi}_{AB}.
\label{DND}
\end{equation}
The non-vanishing components of $T^{\phi}_{AB}$	are:
\begin{eqnarray}
T^{\phi}_{(0)(0)}&=&\beta\Bigg[\frac{1}{2}\left( {\frac {\partial \phi }{\partial t}}  
 \right) ^{2}+\frac{1}{2}{\frac { \left( {\frac {\partial \phi }{\partial t}}
  \right) ^{2} \left(  H \left( r \right) 
 \right) ^{2}}{ \left(  D \left( r \right)  \right) ^{2}}}+\frac{1}{2}
 \left( {\frac {\partial \phi}{\partial r}}  
 \right) ^{2}+\frac{1}{2} \left( {\frac {\partial \phi }{\partial z}}   \right) ^{2}- \frac{H \left( r \right)}{\left(  D \left( r \right)  \right) ^{2} }  \left( {\frac {\partial \phi }{\partial t}}  \right) \bigg({\frac {\partial \phi }{\partial \theta}}\bigg)\nonumber\\
 &+&\frac{1}{2(D(r))^2}\left( {\frac {\partial \phi }{\partial \theta}}  
 \right) ^{2}\Bigg] ;
\nonumber\\
T^{\phi}_{(0)(1)}&=&-\beta\left( {\frac {\partial \phi }{\partial t}} 
 \right) {\frac {\partial \phi }{\partial r}};\nonumber\\
 T^{\phi}_{(0)(2)}&=&\beta\Bigg[\frac { H \left( r \right)}{D \left( r \right)}  \left( \frac {\partial \phi}{\partial 
t}\right)^{2}-\frac{1}{D(r)}\left( {\frac {\partial \phi }{\partial t}}  \right) \bigg({\frac {\partial \phi }{\partial \theta}}\bigg)\Bigg];
\nonumber\\
T^{\phi}_{(0)(3)}&=&-\beta\left( {\frac {\partial \phi }{\partial t}}  
 \right) {\frac {\partial \phi }{\partial z}};\nonumber\\
 T^{\phi}_{(1)(1)}&=&\beta\Bigg[\frac{1}{2} \left( {\frac {\partial \phi }{\partial r}} 
 \right) ^{2}-\frac{1}{2}{\frac { \left( {\frac {\partial \phi }{\partial t}}  \right) ^{2} \left(  H \left( r \right) 
 \right) ^{2}}{ \left( D \left( r \right)  \right) ^{2}}}+\frac{1}{2}
 \left( {\frac {\partial \phi }{\partial t}} 
 \right) ^{2}-\frac{1}{2}\left( {\frac {\partial \phi }{\partial z}}  \right) ^{2}+\frac{H \left( r \right)}{\left(  D \left( r \right)  \right) ^{2} }  \left( {\frac {\partial \phi }{\partial t}}  \right) \bigg({\frac {\partial \phi }{\partial \theta}}\bigg)\nonumber\\
 &-&\frac{1}{2(D(r))^2}\left( {\frac {\partial \phi }{\partial \theta}}  
 \right) ^{2}\Bigg];\nonumber\\
T^{\phi}_{(1)(2)}&=&\beta\Bigg[-\frac { H \left( r \right)}{D \left( r \right)}  \left( \frac {\partial \phi}{\partial 
r}\right)\left( {\frac {\partial \phi }{\partial t}}  \right)+\frac{1}{D(r)}\left( {\frac {\partial \phi }{\partial r}}  \right) \bigg({\frac {\partial \phi }{\partial \theta}}\bigg)\Bigg];\nonumber\\
T^{\phi}_{(1)(3)}&=&\beta\left( {\frac {\partial \phi }{\partial r}} 
 \right) {\frac {\partial \phi }{\partial z}} ;\nonumber\\
T^{\phi}_{(2)(2)}&=&\beta\Bigg[\frac{1}{2}\left( {\frac {\partial \phi }{\partial t}}  
 \right) ^{2}+\frac{1}{2}{\frac { \left( {\frac {\partial \phi }{\partial t}}
  \right) ^{2} \left(  H \left( r \right) 
 \right) ^{2}}{ \left(  D \left( r \right)  \right) ^{2}}}-\frac{1}{2}
 \left( {\frac {\partial \phi}{\partial r}}  
 \right) ^{2}-\frac{1}{2} \left( {\frac {\partial \phi }{\partial z}}   \right) ^{2}- \frac{H \left( r \right)}{\left(  D \left( r \right)  \right) ^{2} }  \left( {\frac {\partial \phi }{\partial t}}  \right) \bigg({\frac {\partial \phi }{\partial \theta}}\bigg)\nonumber\\
 &+&\frac{1}{2(D(r))^2}\left( {\frac {\partial \phi }{\partial \theta}}  
 \right) ^{2}\Bigg] ;\nonumber\\
T^{\phi}_{(2)(3)}&=&\beta\Bigg[-\frac { H \left( r \right)}{D \left( r \right)}  \left( \frac {\partial \phi}{\partial 
z}\right)\left( {\frac {\partial \phi }{\partial t}}  \right)+\frac{1}{D(r)}\left( {\frac {\partial \phi }{\partial z}}  \right) \bigg({\frac {\partial \phi }{\partial \theta}}\bigg)\Bigg];\nonumber
\end{eqnarray}
\begin{eqnarray}
T^{\phi}_{(3)(3)}&=&\beta\Bigg[\frac{1}{2}\left( {\frac {\partial \phi }{\partial t}}  
 \right) ^{2}-\frac{1}{2}{\frac { \left( {\frac {\partial \phi }{\partial t}}
  \right) ^{2} \left(  H \left( r \right) 
 \right) ^{2}}{ \left(  D \left( r \right)  \right) ^{2}}}-\frac{1}{2}
 \left( {\frac {\partial \phi}{\partial r}}  
 \right) ^{2}+\frac{1}{2} \left( {\frac {\partial \phi }{\partial z}}   \right) ^{2}+ \frac{H \left( r \right)}{\left(  D \left( r \right)  \right) ^{2} }  \left( {\frac {\partial \phi }{\partial t}}  \right) \bigg({\frac {\partial \phi }{\partial \theta}}\bigg)\nonumber\\
 &-&\frac{1}{2(D(r))^2}\left( {\frac {\partial \phi }{\partial \theta}}  
 \right) ^{2}\Bigg] .
 \label{T}
\end{eqnarray}
where $C_{AB}$ is given by Eq. (\ref{CT}). Since we want to obtain nontrivial solutions, i.e, those ones corresponding to a some non-zero $C_{AB}$, thus the Eqs. (\ref{CT},\ref{KGd},\ref{DND},\ref{T}) lead to
\begin{equation}
\phi(t,r,\theta,z)=b(z-z_0),
\label{phid}
\end{equation}
note that the solution given by Eq. (\ref{phid}) for the CS field produces  $T^{\phi}_{AB}$ and $C_{AB}$, constants and diagonals, thus satisfying the requirement that $R_{AB}$, $T_{AB}$ and the cosmological constant term are diagonals and constants in tetrad frame (\ref{TB}). It is noticed that in contrast to the non-dynamical framework, where there exist a larger arbitrariness on the choice of the CS scalar field, in the dynamical framework the form of the CS scalar field is more restricted but yet arbitrary due to the $b$ parameter.  

Hence, the Eqs. (\ref{KGd}) become
 \begin{eqnarray}
 (0,0): \quad 2\omega^2+2k(4\omega^2-m^2)+\Lambda-\frac{1}{2}e^{2}-\frac{1}{2}\rho-\frac{3}{2}p &=& 0,  \label{eq00}\nonumber\\
  (1,1)=(2,2): \quad 2\omega^2-m^2+k(4\omega^2- m^2)-\Lambda+\frac{1}{2}p-\frac{1}{2}e^{2}-\frac{1}{2}\rho &=& 0, \\
\label{eq11}  
   (3,3): \quad -\Lambda-(\beta b)^2+\frac{1}{2}p+\frac{1}{2}e^{2}-s^2-\frac{1}{2}\rho &=& 0 .\nonumber
  \label{eqlamb}
 \end{eqnarray}
We define one new parameter, $k = b\omega$, that is, the product between the amplitude of the Chern-Simons field and vorticity of matter, and its physical interpretation is related to the coupling between the gradient of the Chern-Simons field and rotation vector of the G\"{o}del-type universe, both along the preferred direction \cite{PP}.
It is important to note that the $(3,3)$ component of modified field equations is independent of the metric parameters $(m,\omega)$.
 
Naturally, one realizes the similarity with the non-dynamical case \cite{PP}, which is replayed by taking $\beta=0$ in Eq. (\ref{eq00}), but our proposal in this paper is to work in dynamical framework, therefore $\beta\neq 0$. The dynamics of CS field emerges from Eq. (\ref{eq00}) as a contribution of energy momentum tensor of CS field, that is proportional to $b^2$. Besides, under taking $b=0$ in Eq.(\ref{eq00}) one recovers the solution of General Relativity \cite{Reb} as expected in dynamical framework as well.
 
 Before treating a general case, we consider the case without matter but with the cosmological constant, so the modified equations reduce to
\begin{eqnarray}
\label{sys1a}
&&2\omega^2+8k\omega^2-2k m^2-\Lambda=0,\nonumber\\
&&2\omega^2-m^2+4k\omega^2-k m^2+\Lambda=0,\nonumber\\
&&\Lambda=-(\beta b)^2,
\end{eqnarray} 
in order to obtain the solutions for the previous equation, then let us divide them in two cases: the first one is valid for $k\neq -1/3$, which can be obtained adding up the first and second equations of (\ref{sys1a}) after eliminating $\Lambda$ by means of third ones, then, it remains 
\begin{eqnarray}
&&2\omega^2+8k\omega^2-2k m^2+(\beta b)^2=0,\nonumber\\
&&2\omega^2-m^2+4k\omega^2-k m^2-(\beta b)^2=0,
\end{eqnarray}
a priori the system of equation seems reducing to one cubic equation in $\omega$, but if we sum
both equations, we have a simple shape
\begin{equation}
(4\omega^2-m^2)(1+3k)=0,
\end{equation}
whose solution is $m^2=4\omega^2$ for $k\neq-1/3$, as we require solutions of the type $\omega=\omega(k)$ expressed in terms of $k$. Clearly this solution excludes any contribution coming from Cotton tensor, as can be easily verified from Eq. (\ref{CT}) and, thus, it does not present a new situation because reduces to GR. On the other hand, the case $k=-1/3$ is  most interesting solution
 \begin{eqnarray}
\label{L}& \Lambda=-(\beta b)^2,\nonumber\\
\label{b}& b\omega =k=-\frac{1}{3},\nonumber\\
& m^2=\omega^2+\frac{3}{2}(\beta b)^2,
\label{m2}
 \end{eqnarray}
where, for the sake of the convenience, we choose expressing the metric parameters $m^2$ and $\omega$ in terms of CS parameter $b$. Note that if $\Lambda=0$, we have two possibilities the first one is $b=0$ and the  Eqs. (\ref{sys1a}) are trivially solved, thus, it resulting in $m=\omega=0$, the second one we have $\beta=0$ that corresponds to the non-dynamical case. Consequently, the way out to obtain vacuum-solutions in the dynamical framework is required by $\Lambda\neq 0$ which corresponds to $b\neq 0$ and $\beta\neq 0$. In this case,  the solution given by Eq. (\ref{m2}) demands that the cosmological constant is always negative. This solution presents one similarity with the non-dynamical vacuum solution, that is, the coefficient $k=-1/3$ is equal in both, however the Eq.(\ref{m2}) shows  new possibilities in relation to  non-dynamical solution that presents just one geometry, $m^2=\omega^2$, i.e.,  the dynamics of CS field interferes directly on the possible  G\"{o}del-type solutions that satisfy, $m^2>\omega^2$, thus solutions inside the completely causal hyperbolic class, i.e, without CTCs  are possible contrary to the non-dynamical framework. From Eq.(\ref{m2}), these causal solutions are obtained through the requirement $\beta b^2\geq\sqrt{2}/3$ that implies $m^2\geq 4\omega^2$.
 
  In the case with matter sources it is convenient to rewrite the  Eqs.(\ref{eq00}) in terms of the parameter $k$ as shown below 
 \begin{eqnarray}
\omega^{4}&-&\bigg(\frac{a_{\omega}k+b_{\omega}}{3k+1}\bigg)\omega^2-\frac{1}{2}\beta k)^2=0,\label{eq:cs_w2}\nonumber\\
m^{2}&=&\frac{a_{m}k+b_{m}}{3k+1}+\frac{2(\beta k)^2}{\omega^2},
\end{eqnarray}
where the coefficients are defined by 
 \begin{eqnarray}
2\,a_{\omega} &=&  \rho+p+3\,s^{2}-2\,e^{2},\nonumber\\
2\,b_{\omega} &=&  \rho+p+s^{2},\nonumber\\
a_{m} &=&  2\,(\rho+p+3\,s^{2}-2\,e^{2}),\nonumber\\
b_{m} &=&  \rho+p+2\,s^{2}-e^{2},\label{eq:cs_bm2}
\end{eqnarray}
are valid for $k\neq-\frac{1}{3}$ and determined exclusively by the matter content.  One can check that the general relations among the coefficients of $\omega^2$ and $m^2$ are the same as of the non-dynamical case \cite{PP}.
 
 From Eq. (\ref{eq:cs_w2}) we obtain 
 \begin{equation}
 \omega^2(k)=\frac{1}{2}\,\Bigg[\,\bigg(\frac{a_{\omega}k+b_{\omega}}{3k+1}\bigg)+\,\sqrt {{\bigg(\frac{a_{\omega}k+b_{\omega}}{3k+1}\bigg)}^{2}+2\,({\beta k})^{2}}\Bigg],
 \label{eq:cs_w21}
 \end{equation}
 it is observed that $\omega^2\geq 0$ for all $k$, except for $k=-1/3$. Expressing $\omega^2$ by means of Eq. (\ref{eq:cs_w2}) we get the solution for $m^{2}$ in terms of the parameter $k$ 
\begin{equation}
m^{2}(k)=\frac{b_{m}-4b_{\omega}}{3k+1}+2\Bigg(\frac{a_{\omega}k+b_{\omega}}{3k+1}+\sqrt{\bigg(\frac{a_{\omega}k+b_{\omega}}{3k+1}\bigg)^2+2(\beta k)^2}\Bigg),
\label{eq:cs_m21}
\end{equation}
note that the coefficients (\ref{eq:cs_bm2}) together with the $k$ parameter determine the possible signals of $m^2$ and, consequently, the possible classes of the G\"{o}del-type metrics. By combining the Eqs. (\ref{eq:cs_w2}) we obtain the same relation as in the non-dynamical case
\begin{equation}
m^{2}-4\,\omega^{2}=-\frac{\rho+p+e^{2}}{3k+1}.\label{eq:cs_causal}
\end{equation}
The sign in this equation determines the causality properties, i.e., for $k<-1/3$ implies $m^2>4\omega^2$, since the solution is causal, whereas  for $k>-1/3$ implies $m^2<4\omega^2$, the solution is non-causal. Furthermore, it is independent of the parameter $b^2$, thus the dynamics of CS field does not interfere in the causality properties which are exclusively determined by the matter sources and $k$ parameter. In contrast to the non-dynamical framework, now $\omega^2$ is real for all $k$, except in $k=-1/3$ that corresponds to the vacuum solution, it is interesting because there will always be no CTCs for some range of $k$  independently of the matter source considered. 

 Additionally,  we obtain the solution for pure scalar field  from Eq.(\ref{eq:cs_causal}), that corresponds to $m^2=4\omega^2$, such solution is identical to the non-dynamical case with pure scalar field, thus the inclusion of the dynamical term does not modify the solution implying that the Cotton tensor is zero, $C_{A B}=0$. The modified field equations (\ref{eq00}) do not reduce to GR field equations  as in the non-dynamical case, because now there is energy-momentum tensor for the CS field. Nonetheless, if we interpret the CS field as one matter field the equations of motion reduce to the Einstein field equations with two scalar fields. 
  
Now, we shall deal with the solutions for the three classes of the G\"{o}del-type metrics: hyperbolic, trigonometric and linear.

The hyperbolic solution is illustrated by taking, for simplicity, only dust as matter source, then the Eq.(\ref{eq:cs_w21}-\ref{eq:cs_m21}) become 
\begin{eqnarray}
\omega^2(k)&=&\frac{1}{4}\,\Bigg[\,\bigg(\frac{\rho k+\rho}{3k+1}\bigg)+\,\sqrt {{\bigg(\frac{\rho k+ \rho}{3k+1}\bigg)}^{2}+8\,({\beta k})^{2}}\Bigg],\nonumber\\
m^{2}(k)&=&\frac{\rho k}{3k+1}+\sqrt{\bigg(\frac{\rho k+\rho}{3k+1}\bigg)^2+8(\beta k)^2},
\end{eqnarray}
on the asymptotic limit of $k$, i.e, $k\rightarrow\pm\infty$, the functions $\omega^2$ and $m^2$ tend to $+\infty$. It is worth mentioning that $m^2$ is always positive for all $k$ and $\rho$, so this solution belongs to the hyperbolic class. 
Similarly to the non-dynamical framework there are causal solutions inside the range $k<-1/3$, namely, now the range of $k$ for causal solutions is greater than in non-dynamical framework \cite{PP}.

\begin{figure}[ht]
\centering
\includegraphics[scale=0.8]{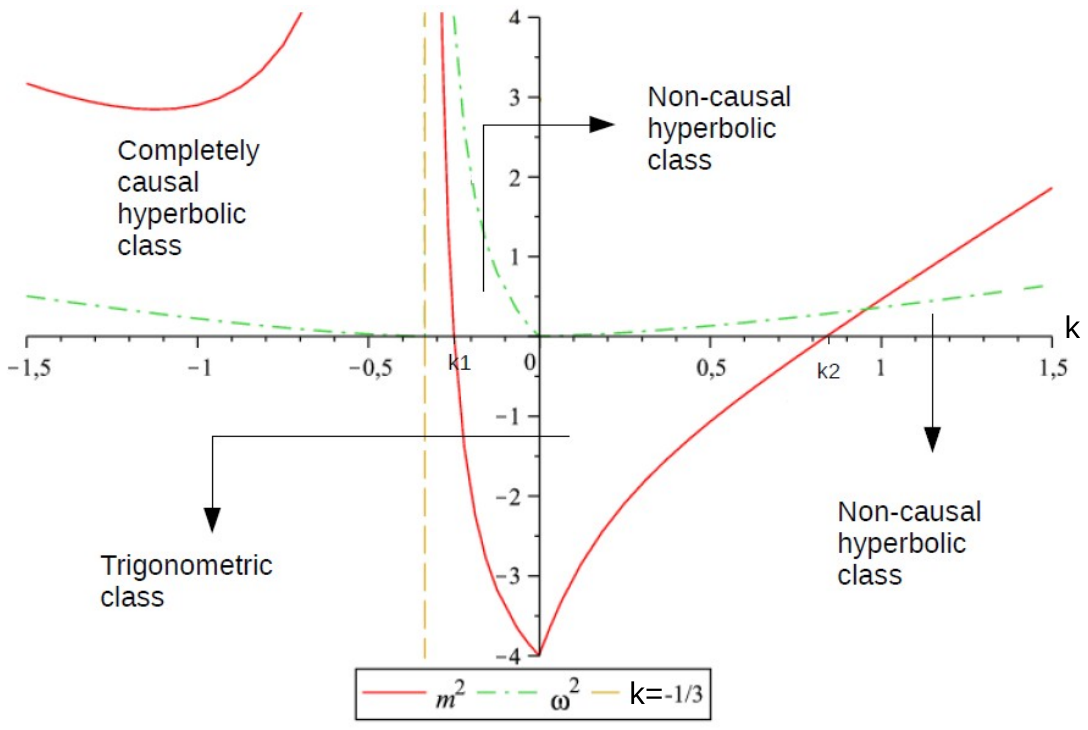}   
\caption{\label{electro}CS solution $\omega^2(k)$ and $m^{2}(k)$ for pure electromagnetic field
 with $e=2$. The yellow dashed line corresponds to $k=-1/3$, where the functions $m^2$ and $\omega^2$ blow up.  }
\end{figure}

All these classes of the G\"{o}del-type metric can be obtained by treating as source just the electromagnetic field, so the Eqs. (\ref{eq:cs_m21},\ref{eq:cs_causal}) reduce to
\begin{eqnarray}
\omega^{2}(k)&=&-\frac{1}{2}\bigg({\frac {{e}^{2}k}{3\,k+1}}\bigg)+\frac{1}{2}\,\sqrt {{\bigg(\frac {{e}^{2}{k
}}{3\,k+1}}\bigg)^2+2\,({\beta k})^{2}},\\
m^{2}(k)&=&-\bigg({\frac {2{e}^{2}k+e^2}{3\,k+1}}\bigg)+2\sqrt {\bigg(\frac {e^{2}
k}{3\,k+1}\bigg)^2+2\,({\beta k})^{2}}.
\label{m22}
\end{eqnarray}
Note that the  roots of Eq. (\ref{m22}) are obtained by solving the resultant equation which is fourth-order one, namely,
\begin{equation}
\beta^2(72k^4+48k^3+8k^2)-4e^4 k-e^4=0.
\end{equation}
despite its analytical solving is possible it is unworkable. Then, to obtain some physical information we solve this equation numerically. For this purpose we shall take $e=2$ and $\beta=1$, thereat we obtain two numerical solutions with three-digit accuracy: $k_{1}=-0,249$ and $k_{2}=0,839$. In the following we analyze  the graphic depicted at Fig. \ref{electro}. Therefore, the three classes of G\"{o}del-type metrics are distinguished by the value of $k$. The first one, the hyperbolic class, occurs within the regions: $k<-1/3$ that corresponds the completely causal region, i.e., without CTCs, $-1/3<k<k_1$ and $k>k_2$ that correspond the non-causal region. The linear class is obtained for $k=k_1$ or $k=k_2$. The region of range between the roots, $k_1<k<k_2$, stands for the trigonometric class. So, it is possible to have a completely causal solution which cannot arise for the presence of the pure electromagnetic field in the non-dynamical case.
 
 \section{Summary}

  We studied the G\"{o}del-type metrics and its consistency within the dynamical  CS modified gravity. We used the same procedure described in \cite{PP}, similarly,  it permitted one more simplification of the calculations. Analogously to the non-dynamical case, we show that when the parameter $b$ is zero, one recovers the GR as expected.  

  We verified that the vacuum case is a solution in the dynamical case, an namely in this case one finds conditions  for $b$ parameter, and hence for the CS field, which permits completely causal G\"{o}del-type metrics, i.e. there is no CTCs. Particularly, this result is remarkable because even without matter the dynamics of CS field permits causal solutions which cannot take place in the non-dynamical framework.
  
As for matter sources, our main result is obtaining causal solutions for any matter source. It is worth to call the attention that the parameters of the amplitude of the Chern-Simons field $b$ and vorticity of matter $\omega$ couple  in modified field equations, as a consequence we define a new parameter $k=b\omega$. Therefore, this new framework along to the fact that in the dynamical framework the restriction of $\omega^2$ to be real is valid for all values of the parameter $k$, except $k=-1/3$, it allowed such causal solutions. For example, the pure electromagnetic field in the non-dynamical framework does not present completely causal solutions, see \cite{PP}, however as we shown there are CTC-free solutions for all matter sources, in particular, for pure electromagnetic field.
  
\textbf{Acknowledgments.} This work was partially supported by Conselho
Nacional de Desenvolvimento Científico e Tecnológico (CNPq). The work
by A. Yu. P. has been supported by the CNPq project No. 303783/2015-0.

\end{document}